# Spectra-Orthogonal Optical Anisotropy in Wafer-Scale Molecular Crystal Monolayers

*Tomojit Chowdhury,[1,2] Fauzia Mujid,[1] Zehra Naqvi,[3] Ariana Ray,[4] Ce Liang,[3] David A. Muller,[4] Nathan P. Guisinger,[5] and Jiwoong Park[1,2,3],\**

[1]Department of Chemistry, University of Chicago, Chicago, Illinois 60637, United States

[2]The James Frank Institute, University of Chicago, Chicago, Illinois 60637, United States

[3]Pritzker School of Molecular Engineering, University of Chicago, Chicago, Illinois 60637, United States

[4]Department of Applied and Engineering Physics, Cornell University, Ithaca, New York 14853, United States

[5]Center for Nanoscale Materials, Argonne National Laboratory, Argonne, Illinois 60439, United States

**ABSTRACT**

Controlling the spectral and polarization responses of two-dimensional (2D) crystals is vital for developing ultrathin platforms for compact optoelectronic devices. However, independently tuning optical anisotropy and spectral response remains challenging in conventional semiconductors due to the intertwined nature of their lattice and electronic structures. Here, we report spectra-orthogonal optical anisotropy—where polarization anisotropy is tuned independently of spectral response—in wafer-scale, one-atom-thick 2D molecular crystal (2DMC) monolayers synthesized on monolayer transition metal dichalcogenide (TMD) crystals. Utilizing the concomitant spectral consistency and structural tunability of perylene derivatives, we demonstrate tunable optical polarization anisotropy in 2DMCs with similar spectral profiles, as confirmed by room-temperature scanning tunneling microscopy and cross-polarized reflectance microscopy. Additional angle-dependent analysis of the single- and polycrystalline molecular domains reveals an epitaxial relationship between the 2DMC and the TMD. Our results establish a scalable, molecule-based 2D crystalline platform for unique and tunable functionalities unattainable in covalent 2D solids.

Thin-film, solid-state semiconductor materials have driven major advances in optoelectronic technologies,[1,2] but unlocking their full potential requires arbitrary control over their spectral and polarization responses . For this purpose, 2D semiconductors, such as monolayer TMDs,[3] provide an ultrathin materials platform with myriad exciting spectral signatures relevant to optoelectronic systems and devices.[4–8] However, their covalently bonded, isotropic (typically honeycomb) lattices inherently intertwine their spectral and polarization properties,[9] making independent regulation of these features challenging within a single material. On the other hand, structurally anisotropic 2D semiconductors (e.g., black phosphorus) demonstrate tunable optical polarization at a specific energy,[10,11] but its rigid lattice, susceptible to rapid degradation under ambient conditions,[12] limits its practicality as a scalable platform for harnessing anisotropic behavior. In contrast, 2DMCs—formed through the planar arrangement of non-covalently linked discrete molecules containing chromophores and functional groups[13–15]—may offer the potential for spectra-orthogonal tuning of optical anisotropy. This is enabled by (1) the chromophore, which determines the crystal's spectral peak position, and (2) functional groups on the chromophore, which modify the 2DMC lattice geometry and, in turn, may impart differing optical polarization responses. Despite significant progress in controlling the properties of 2DMCs,[16–26] however, the realization of independent control on the spectral and optical polarization is significantly lacking, mainly due to a lack of wafer-scale homogeneous, highly crystalline materials.

Here, we demonstrate that monolayer perylene-based 2DMCs are promising candidates for tuning of optical anisotropy independent of their spectral response through a combined synthesis and characterization investigation. The wafer-scale 2DMC monolayers used in this study were grown by vapor phase synthesis on monolayer TMD surfaces. A suite of high resolution scanning probe microscopy experiments reveal that the 2DMC structures—including lattice size, shape, and density—can be systematically tuned with specific changes to the functional groups on the chromophore. We show that this structural variation is directly correlated with 2DMC's optical polarization, measured at similar spectral positions using cross-polarized microscopy under ambient conditions. Further statistical analysis of angle-dependent optical data and electron diffraction experiments reveals a strong alignment between the lattice orientations of molecular domains and the TMD, facilitating the synthesis of a new class of structurally compatible bilayer crystals.

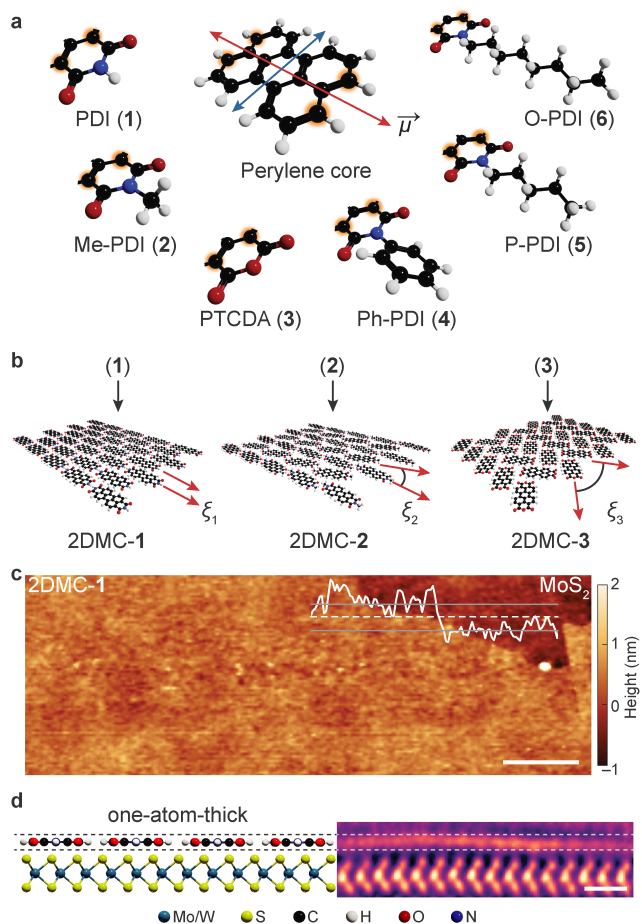

**Figure 1. Compositional and structural features of 2DMC building blocks.** (a) Ball-and-stick models of the perylene core, illustrating the primary and secondary dipole vectors, represented by red and blue double sided arrows, respectively, along with six peri-substituted functional groups (labelled 1–6). (b) Energy-optimized structures of three monolayer 2DMCs derived from different perylene derivatives (1–3), highlighting variations in the spatial orientation of molecular dipoles (red arrows) and the corresponding misorientation angles. (c) AFM topographic map of partially grown 2DMC-1 on monolayer $MoS_2$ surface. An AFM height profile is overlaid, taken along the dashed white line, highlighting the step height by grey lines. Scale bar: 500 nm. (d) Ball-and-stick model (side-view) of a one-atom-thick 2DMC on a TMD monolayer (left). A representative cross-sectional iDPC ac-STEM image of a portion of 2DMC-1 grown on $MoS_2$ monolayer highlights the 2DMC thickness, marked with dashed grey lines. Scale bar: 1 nm.

**Figure 1** introduces the 2DMCs synthesized and studied in this work. These 2DMCs are based on molecular units (1–6; **Figure 1a**) that share a two-fold symmetric and structurally anisotropic perylene core but differ only in their peri-functional groups. The absorption and emission of monochromatic light by these molecules are polarized along their long axis ($\vec{\mu}$) in the visible range,[27,28] making them optically anisotropic building blocks. The polarization properties of a molecular crystal can thus be controlled by the planar crystalline arrangement of these molecules or dipoles, as illustrated in **Figure 1b**. This arrangement is realized for three functional groups——NH in 2DMC-1, –NCH$_3$ in 2DMC-2, and –O– in 2DMC-3 (**Figure 1b**)—with specific intermolecular interactions, mainly in-plane hydrogen bonds.[29] These interactions generate distinct dipolar alignments within the 2DMC, leading to unique 2D geometries with misorientation angles ($\xi_1$, $\xi_2$, and $\xi_3$; **Figure 1b**) defined by the long axes of the molecules in adjacent rows.

We synthesized 2DMCs atop as-grown monolayer TMDs (MoS$_2$ or WS$_2$) via self-limited physical vapor deposition (PVD) (see Methods and Table S1 for details of synthesis). The morphology, topography, and thickness of the resulting 2DMCs were verified using optical microscopy (see

Figure S1) and atomic-resolution scanning probe measurements (see Methods for details). Fast-scanning atomic force microscopy (AFM) revealed a step height of less than 5 Å at the interface and high spatial uniformity of 2DMC-1 atop monolayer MoS$_2$ (**Figure 1c**). To further investigate the interface structure and quality with atomic resolution, a cross-sectional segment of 2DMC-1 on MoS$_2$ was prepared using focused ion beam (FIB) milling and subsequently analyzed via aberration-corrected scanning transmission electron microscopy (ac-STEM). We employed integrated differential phase contrast (iDPC) STEM imaging, which confirmed a one-atom-thick (~4 Å; **Figure 1d**) 2DMC layer forming an atomically abrupt interface with the three-atom-thick MoS$_2$ monolayer. The interlayer (center-to-center) distance was measured to be ~5 Å, consistent with the van der Waals interaction regime.[30] Notably, iDPC, which represents the sample's projected electrostatic potential, contrasts weakly with atomic number.[31–33] This makes it particularly effective for simultaneously detecting the much lighter 2DMC layer—predominantly a hydrocarbon network—and the heavier TMD layer with atomic resolution. Compared to other STEM modes, such as high-angle annular dark-field imaging, iDPC also provides better dose efficiency and lower noise, making it ideal for visualizing electron beam-sensitive molecular monolayers.

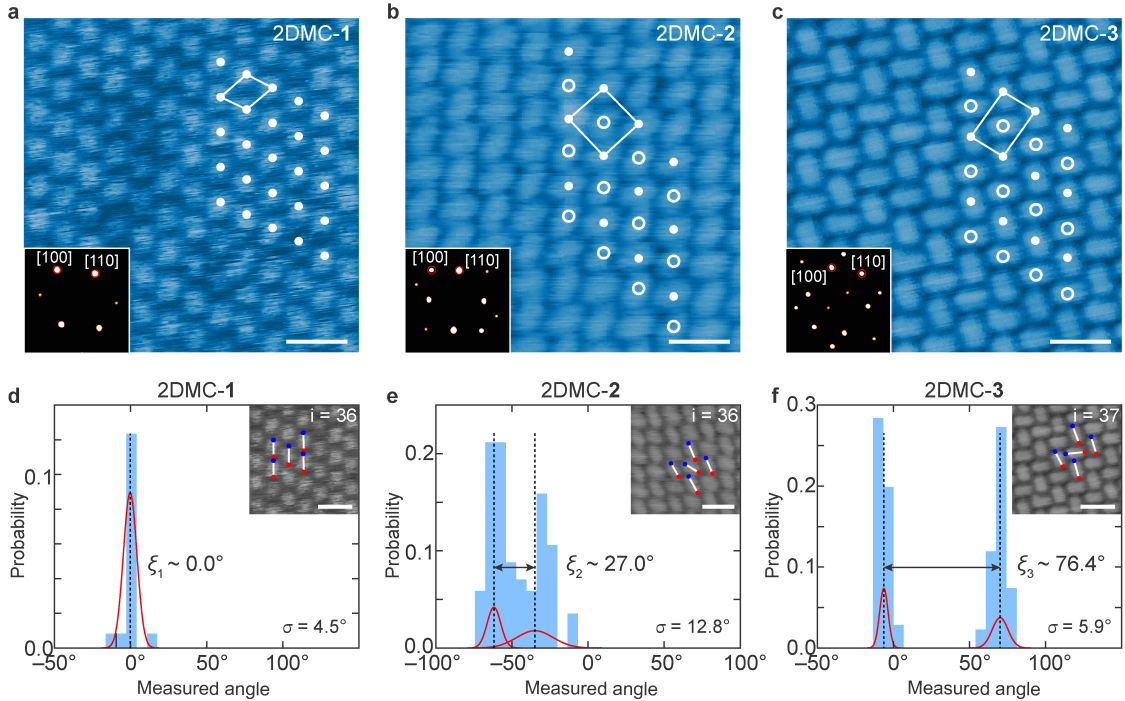

**Figure 2. 2DMC lattice structural tuning and characterization.** (a–c) Room-temperature UHV-STM topographic images of 2DMC-1 (a; $V = -800$ mV, $I = 50$ pA), 2DMC-2 (b; $V = +2$ V, $I = 100$ pA), and 2DMC-3 (c; $V = -1$ V, $I = 100$ pA), grown on monolayer MoS$_2$ and transferred onto Au(111)/mica substrates. *Insets:* FFT images of the molecular lattices, highlighting the (100) and (110) Bragg peaks. Solid and open white dots denote molecular lattice points, with molecular unit cells outlined by white boxes. Scale bars: 2 nm. (d–f) Histograms of the angular distributions for the long axes of molecular units in each 2DMC, derived from pixel-based analysis of STM images. Red curves represent first-order Gaussian fits to the histogram data, from which mean angles (black dotted lines) and standard deviations ($\sigma$) are extracted. Misorientation angles ($\xi_1$, $\xi_2$, and $\xi_3$) are calculated as the absolute differences between the mean angles (double-sided arrows), with root-sum-square values of $\sigma$ for 2DMC-2 (e) and 2DMC-3 (f), and as the mean value $\pm\sigma$ for 2DMC-1 (d). *Insets:* Grey-scale STM images showing the endpoints of molecular units, with red and blue dots overlaid and white lines marking the long axes of the molecular units (i = number of total molecular units analyzed in each image). Scale bars: 2 nm.

Ultrahigh vacuum (UHV) scanning tunneling microscopy (STM) experiments were conducted to differentiate between various 2DMC crystal structures, with three representative examples shown and discussed in **Figure 2** (see Methods for further details). For STM imaging, large-area, continuous 2DMC films grown on $MoS_2$ monolayers were first water-transferred onto conductive Au(111) surfaces. This polymer-free transfer step ensured clean surfaces free of residues and impurities, allowing for high-quality imaging. **Figures 2a–c** present plan-view STM images of 2DMCs, showing their distinct lattice structures—brick-wall (2DMC-1), canted (2DMC-2), and herringbone (2DMC-3) configurations.[19,20,34] These lattices consist of molecular unit cells with lattice constants ranging from 1 to 2 nm, which are 3 to 6 times larger than the lattice constant of $MoS_2$.[35] The measured molecular lattice parameters are detailed in Table S2 (see also Figure S2). While all three 2DMCs exhibit nearly hexagonal 2D lattice arrangements—confirmed by the Fourier transforms of the STM images (insets; **Figures 2a–c**)—their real-space lattice structures and intermolecular spacings differ significantly, as highlighted by the overlaying arrays of white dots.

To measure the relative angular orientation of 2DMC lattices, we analyzed the orientation of individual molecules using STM images through a pixel-based method. The process involved: (1) extracting the molecule's center position and determining the angular distribution of its pixels, and (2) using this pixel-distribution to identify the molecule's long axis (see Supporting Information for further details of the statistical analysis). **Figures 2d–f** show the long-axis distributions for the 2DMCs, with insets marking the long axis using lines with red and blue endpoints. For 2DMC-1 (**Figure 2d**), we observed a single narrow distribution with a standard deviation of 4.5°, indicating highly aligned molecules with negligible misorientation ($\xi_1$ ~0°). In contrast, 2DMC-2 and 2DMC-3 displayed two distinct peaks in their distributions, each corresponding to a subset (nearly half) of molecules with different orientations. The misorientation angles, $\xi_2 = 27.0°$ and $\xi_3 = 76.4°$, were measured from the center-to-center distances (**Figures 2e, f**), with root-sum-square uncertainties of 12.8° and 5.9°, respectively.

These results align well with our hypothesis that functional groups markedly influence 2DMC lattice structures (**Figure 1b**), revealing a key difference for 2DMC-1 compared to previous studies.[19,34] Earlier work on PDI molecular lattices reported ~10–20° misorientation, whereas we observe near-perfect alignment in PDI-containing 2DMC-1. This discrepancy likely stems from the different TMD substrates used: while previous studies utilized bulk $MoS_2$ for STM and diffraction measurements,[19,34] we employed monolayer $MoS_2$ and $WS_2$. Changes in the TMD substrate—monolayer versus bulk—affect PDI–TMD interactions, as also seen in the molecular density. For 2DMC-1, the molecular density is 10% higher than that of 2DMC-3 (see Table S1), compared to a ~1% difference reported previously.[19,34] This suggests that the PDI molecules in our 2DMC-1 adopt a much denser, more distinct lattice structure than observed in previous studies.

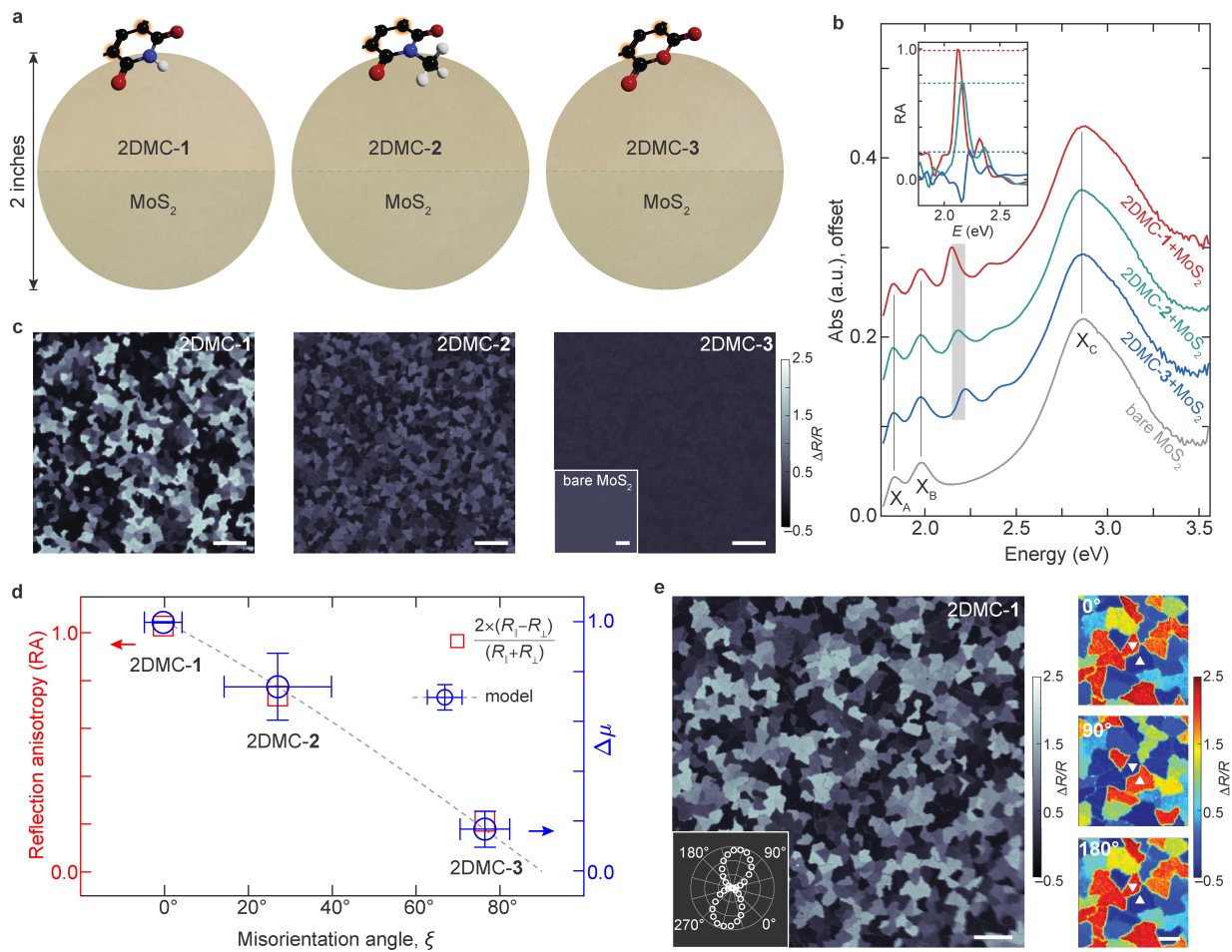

**Figure 3. Spectra-orthogonal polarization tuning in large-scale 2DMCs.** (a) Color photographs of three as-grown 2DMC films (1–3) on MoS$_2$ (upper half) and bare MoS$_2$ (lower half) on 2" fused silica wafers. (b) Unpolarized UV-vis absorption spectra of the 2DMCs, with the primary molecular absorption region highlighted by a grey shaded box (2.18 eV ± 30 meV). Bare MoS$_2$ spectrum is shown side-by-side to identify its excitonic peaks (labelled as X$_A$, X$_B$, and X$_C$, marked with black lines). *Inset:* Reflection anisotropy (RA) plot for the 2DMCs, with different RA ranges indicated by dashed lines. (c) $\Delta R/R$ maps of 2DMCs (1–3; $E \sim 2.2$ eV) on MoS$_2$ and bare MoS$_2$ ($E$ = 1.89 eV) in the inset. Scale bars: 20 µm. (d) Plot of measured RA values and the vector model $\Delta\mu$, as a function of $\xi$ for the 2DMCs (1–3). (e) Grey-scale $\Delta R/R$ map of 2DMC-1 ($E$ = 2.15 eV) (left). Scale bar: 20 µm. Zoomed-in false-colored maps (right) (at 0°, 90°, and 180°) marking two adjacent single crystal domains by upright and inverted triangular markers, revealing a polar plot from one of the domains. Scale bar: 10 µm.

The precise synthetic control of 2DMC lattices, confirmed through direct imaging of molecular unit cells, enables us to investigate the optical spectra and polarization response of various 2DMCs, as shown in **Figure 3**. **Figure 3a** presents unpolarized color photographs, taken with a phone camera, of three representative 2DMCs—2DMC-1, 2DMC-2, and 2DMC-3—grown on $MoS_2$ monolayers on 2" fused silica wafers (top halves), with bare $MoS_2$ monolayers (bottom halves) for comparison. In addition, our method can produce more than half-a-dozen different wafer-scale 2DMCs with different molecular units (see Figure S3 and Figure S4) and on surfaces beyond TMDs, like graphene and hexagonal boron nitride (see Figure S5), emphasizing the generality of our synthesis method. All 2DMCs exhibit wafer-scale homogeneity with uniform color and contrast, indicating a consistent optical absorption profile attributed to the shared perylene core. **Figure 3b** displays room-temperature UV-vis absorption spectra of the three 2DMCs on $MoS_2$, alongside bare $MoS_2$ on transparent silica substrates. The spectra reveal characteristic absorption peaks near 2.2 eV (± 30 meV) primarily defined by the perylene core[27] (highlighted by the shaded region), along with the common $MoS_2$ excitonic peaks[36,37] ($X_A$, $X_B$, and $X_C$ marked by dotted lines). Room-temperature Raman scattering spectra further confirm the formation of molecular crystals, showing aromatic vibrational fingerprints between 1200 and 1700 cm$^{-1}$ (Figure S6).[38]

The optical polarization response of the 2DMCs was measured using cross-polarized reflectance microscopy. Here, a small angle ($\delta$ ~3°) from perfect cross-polarization (90°) was introduced between the polarizer and analyzer to enhance the signal-to-noise ratio[39] (see Methods and Figure S7 for experimental details). Prior to these measurements, as-grown 2DMCs (on $MoS_2$) were transferred onto transparent (fused silica) substrates to eliminate internal reflections. **Figure 3c** shows wide-field differential reflectance ($\Delta R/R$) maps of the three 2DMCs near the absorption maxima (at $E$ ~2.2 eV). Among these samples, the contrast between the brightest and darkest domains is observed to be highest in 2DMC-1, lowest in 2DMC-3, and intermediate in 2DMC-2, whereas the MoS$_2$ layer alone exhibits uniform reflectance due to its isotropic honeycomb lattice structure (inset; **Figure 3c**). Polarized reflectance spectra for all three 2DMCs show peaks near 2.2 eV (Figure S8). The maximum reflection anisotropy (RA), calculated using the formula $2 \times (R_\parallel - R_\perp)/(R_\parallel + R_\perp)$, is observed in 2DMC-1 (RA ~1), while the minimum in 2DMC-3 (RA ~0.2), where $R_\parallel$ and $R_\perp$ are the reflected intensities recorded from a single crystalline 2DMC domain oriented parallel and perpendicular, respectively, to the incident linear polarization direction (see inset, **Figure 3b**).

To analyze the polarization response in different 2DMCs, we plotted RA as a function of $\xi$, revealing a proportional relationship, as shown in **Figure 3d**. This relationship aligns with an empirical dipole vector model ($\Delta\mu$; see Figure S9 and Supporting Discussion for details of the vector model), which captures two key features: (1) the absorption and reflectance spectra share similar frequency profiles, with the strongest polarization dependence near molecular absorption peaks, indicating that the anisotropic perylene core dictates the peak positions for all 2DMCs; and (2) reflected intensities vary systematically with the functional groups on the perylene core, leading to different $\xi$ values and, consequently, distinct polarization responses near the same photon energy ($E$ ~2.2 eV). Together, these results establish $\xi$ as a synthetically tunable parameter for realizing spectra-orthogonal control of optical polarization responses in 2DMCs.

This systematic polarization control enables detailed analysis of 2DMC domain orientations through angle-dependent reflectance measurements. **Figure 3e** shows greyscale (left) and false-colored (right) $\Delta R/R$ maps of polycrystalline domains of 2DMC-1 on monolayer MoS$_2$, revealing a characteristic two-fold symmetric polar plot (inset, **Figure 3e**). Notably, the domain sizes of the 2DMC-1 closely match those of the underlying MoS$_2$, resulting in uniform angular distributions (see Figure S10). This observation suggests an intricate substrate-crystal interplay that may dictate specific molecular domain orientations, as discussed below.

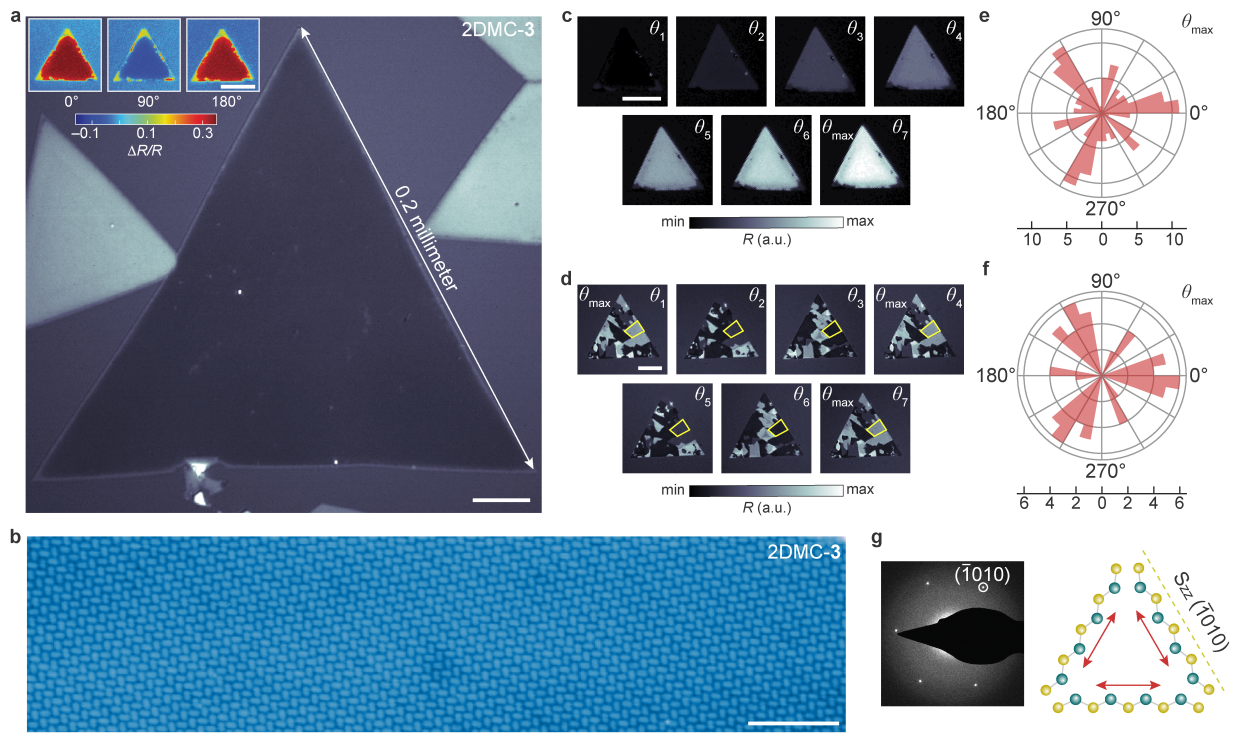

**Figure 4. Epitaxial relationship between 2DMC and TMD monolayers.** (a) Polarized reflectance image of single crystal 2DMC-3 domains ($E = 2.19$ eV) on monolayer WS$_2$ single crystals. *Inset:* Angle ($\theta$)-dependent cross-polarized $\Delta R/R$ maps of 2DMC-1 domains ($E = 2.15$ eV) on WS$_2$ single crystals. Scale bars: 20 µm. (b) Large field-of-view STM topographic image of a single crystal 2DMC-3 domain on a monolayer MoS$_2$ substrate ($V = -1$ V, $I = 100$ pA). Scale bar: 10 nm. (c, d) $\theta$-dependent reflectance images ($E = 2.15$ eV) of 2DMC-1 single crystal (c) and polycrystalline (d) domains on WS$_2$ single crystals. Scale bars: 20 µm. $\theta_{max}$ values from 53 single crystal and a 66 polycrystalline 2DMC-1 domains (an example single domain highlighted with yellow boxes) were used to generate the polar histograms in (e) and (f), respectively. (g) SAED image from a 2DMC-1 on WS$_2$, highlighting the FFT peak ($\bar{1}010$) (left). The schematic (right) illustrates the relative orientation of the 2DMC dipole vectors (faded red double-sided arrows) on a TMD crystal with a zigzag sulfur-terminated (Szz) edge (yellow dashed line).

To explore this substrate-crystal interaction and its impact on molecular orientation, we began by synthesizing highly crystalline, large molecular domains directly on TMD single crystals. **Figure 4a** presents a representative reflectance image of 2DMC-3 grown on triangular WS$_2$ single crystals, which reached sizes as large as 0.2 millimeters. These domains demonstrated perfect single-crystalline order extending over 100 nanometers, as evidenced by the absence of imperfections such as molecular vacancies or dislocations, highlighted in the larger field-of-view STM topography image (**Figure 4b**). **Figure 4a** also shows that the reflectance contrast varies significantly between individual islands grown with different orientations, while remaining uniform within each island. Angle ($\theta$)-dependent $\Delta R/R$ maps at the molecular absorption energy ($E = 2.15$ eV; Supporting Video S1) show a variation in contrast with a 2-fold periodicity (see also inset, **Figure 4a**), while remaining uniform or isotropic at the WS$_2$ excitation energy ($E = 2.08$ eV; see Supporting Video S2). Together, these data—combined with **Figure 3e**—demonstrate that the size of 2DMC domains can be controlled solely by synthesis, with large (small) TMD single domains producing correspondingly large (small) 2DMC domains.

Additional $\theta$-dependent reflectance imaging and analysis of single-crystal (**Figure 4c**) and polycrystalline (**Figure 4d**) 2DMC-1 on single-crystal WS$_2$ islands reveal an epitaxial relationship between the molecule and the TMD. From these images, we measured the angles corresponding to maximum reflectance signals ($\theta_{max}$) and generated polar histograms based on over 50 single-crystal domains in each case—single-crystal (**Figure 4e**) and polycrystalline (**Figure 4f**) 2DMC-1 domains. Each histogram exhibits six peaks, with two peaks appearing within a 120° range, forming a quasi-3-fold symmetric pattern. This pattern aligns with the rotational symmetries of the PDI molecule (2-fold) and WS$_2$ (3-fold), suggesting that a strong molecule-TMD interaction leads to a 3-fold (or 6-fold) symmetric polar histogram.[40] Selected area electron diffraction (SAED) confirms an S-zigzag (Szz) termination along the ($\bar{1}010$) plane of the WS$_2$ crystal (**Figure 4g**, left),[41,42] resulting from WS$_2$ synthesis carried out under a significantly higher chalcogen partial pressure (~60 torr) compared to that of the metallic precursor (~10$^{-4}$ torr).[35] Based on our reflectance and SAED data analysis, we propose that the PDI molecules (in 2DMC-1) align along one of the three equivalent zigzag edges of a WS$_2$ single domain, as illustrated schematically in **Figure 4g** (right; red double sided arrows depict primary molecular lattice dipolar alignment). This alignment results in the quasi-3-fold symmetric pattern observed in the polar histograms, providing strong evidence for an epitaxial registry between the PDI and WS$_2$.

In conclusion, we have demonstrated the independent control of optical spectra and polarization anisotropy in atomically thin, perylene-based 2DMCs. This is enabled by the inch-scale synthesis of chemically stable, highly crystalline 2DMCs on monolayer TMD surfaces, as characterized by atomic resolution STEM and STM measurements, which reveal precise synthetic control over 2D lattice geometries. These differing lattice configurations yield distinct polarization responses while maintaining spectral consistency due to the inherent optical properties of the perylene core, as shown through detailed polarized reflectance imaging and spectroscopy performed under ambient conditions. Moreover, we show that the molecular domains epitaxially align with the TMD surface, supporting the formation of a new class of bilayer crystals with a high degree of lattice coherence at the interface. These large-scale bilayer crystals could offer new and tunable optical properties via the careful manipulation of structural (lattice constants and symmetry), chemical (bonding interactions), and electronic (molecular orbitals and band structures) parameters—advancing next generation ultrathin optoelectronic and photonic devices.

Methods

**Wafer-scale synthesis of 2DMCs.** Inch-scale uniform and monolayer TMDs ($MoS_2$, $WS_2$) were synthesized on both $SiO_2$/Si(100) (University Wafer, 300 nm wet thermal oxide, SSP) and fused silica (University Wafer, thickness 500 μm, DSP) wafers through metal-organic chemical vapor deposition (MOCVD) adapted from a previously reported protocol.[35] The MOCVD conditions were optimized to produce high-quality, monolayer TMDs with morphological characteristics suitable for molecular monolayer deposition and wide-field optical imaging. These characteristics include large single-crystal domain size (*D*) for both laterally-stitched, continuous polycrystal $MoS_2$ grains (*D* ~20 μm) and isolated $WS_2$ triangles (*D* ~50–250 μm). To grow 2DMCs, commercially available high purity (~98–99 w%) powdered molecular precursor (Sigma Aldrich, PTCDA, PDI, Me-PDI, P-PDI, O-PDI, Ph-PDI, and THPP) was loaded into a hot-walled quartz tube PVD reactor without further purification. Prior to growth, the as-grown TMD samples were thoroughly rinsed with HPLC grade methanol and isopropyl alcohol, dried under nitrogen ($N_2$) flow, and promptly positioned (within ~2 min) alongside the molecular precursor in the reactor, which was subsequently pumped down. The precise positioning of the substrate relative to the molecular precursor is critical to ensure self-limited monolayer molecular deposition and prevent multilayered or bulk molecular crystal formation. The PVD reactor was then sequentially flushed with ultrahigh purity $N_2$ gas (flow rate: 1000 standard cubic centimeters per minute or sccm) and evacuated to a base pressure of ~$6\times10^{-4}$ torr. Once the base pressure was reached, the temperature of the furnace was quickly ramped to 120 ºC to remove completely the moisture and low-boiling organic impurities under a 200 sccm $N_2$ flow for ~2–3 hours. For certain 2DMC synthesis, multiple full PVD cycles (i.e., time required to raise the temperature of the furnace, maintain that temperature for a specific duration of time, and cool down to room temperature) were performed to achieve complete molecular monolayer coverage.

**Scanning probe microscopy.** Topography and thickness of the as-grown 2DMC samples were determined by state-of-the-art AFM (Cypher ES equipped with blueDrive™) using a fast-scanning high-frequency silicon probe (FS-1500AUD; tip radius 10 nm) in ambient conditions. For STM measurements, as-grown continuous 2DMC samples on TMD were first water-delaminated from the $SiO_2$/Si growth substrates and then transferred on to Au(111)-coated mica substrates (Phasis) for imaging at room temperature. Measurements utilized a commercial variable temperature STM (VT Scienta Omicron) operating at a base pressure of $5\times10^{-11}$ mbar. Imaging utilized electrochemically sharpened tungsten tips. Samples were annealed overnight in a UHV preparation chamber at approximately 100 ºC to drive off mostly water and any physisorbed contamination. The preparation chamber is connected to the analysis chamber, so the samples remained in situ between annealing and imaging.

**Aberration corrected scanning transmission electron microscopy**. Cross-sectional samples were prepared using a Thermo Fisher Helios G4 UX Focused Ion Beam (FIB) with protective carbon and platinum layers deposited via electron and Ga ion beams. The cross-section was cut at a 90° angle and initially milled at 30 kV, with final milling at 5 kV. Samples were imaged using an aberration-corrected Thermo Fisher Scientific Spectra 300 X-FEG STEM at 300 kV, with a probe semi-convergence angle of 30 mrad. Raw iDPC data were collected on an annular detector split into four equal angular segments covering portions of the bright field disk and low-angle scattering region. The iDPC image was formed by subtracting signals from opposite detector segments to create a vector DPC image, which was then integrated in Fourier space. A high-pass filter was applied to remove the slowly varying carbon contamination background.

**Wide-field reflectance microscopy.** Wide-field cross-polarized differential reflectance or reflection contrast ($\Delta R/R$) images and spectra were acquired in our homebuilt hyperspectral microscope. First, a broadband white light (Hamamatsu Xenon lamp, 400–700 nm) was passed through a monochromator and then coupled to a multimode fiber. The fiber-coupled light was passed through a linear polarizer and a 90:10 dichroic beam splitter and shone on to the sample mounted on a rotatable stage with a strain-free 20× objective lens (NA = 0.5). The reflected light was first passed through the beam splitter and then an analyzing polarizer rotated at a small angle ($\delta \sim 3°$) from the cross-polarizer geometry. This nearly orthogonal polarizer configuration significantly improves the signal to noise ratio by suppressing the isotropic response from the substrate ($SiO_2$ or $TMD/SiO_2$) without notably diminishing the anisotropic response from the molecular domain. Reflected signals were detected by an electron-multiplying CCD (EMCCD) camera (Andor iXon+). This imaging technique provides simultaneous assessment of the material's morphology and macroscopic optical anisotropy across a large field-of-view (~0.5×0.5 $mm^2$). The low-power (~0.03 $mW/mm^2$) and rapid (50–100 ms/frame) imaging prevents the soft molecular crystals against beam-induced damage and photobleaching.

## ASSOCIATED CONTENT

**Supporting Information**. The following files are available free of charge.

Description of the vector model and anisotropy (Discussion section), additional experimental data to support generality of synthesis, optical measurement methods, and domain structure analysis (Figures section), and tables (Tables section).

## AUTHOR INFORMATION

**Corresponding Author**

Jiwoong Park — *Department of Chemistry, University of Chicago, Chicago, Illinois 60637, United States; The James Frank Institute, University of Chicago, Chicago, Illinois 60637, United States; Pritzker School of Molecular Engineering, University of Chicago, Chicago, Illinois 60637, United States*; Email: jwpark@uchicago.edu

Author Contributions

T.C., F.M., and J.P. conceived the project. T.C. and F.M. caried out all experiments with help from Z.N. A.R. performed STEM imaging. N.P.G. assisted with STM imaging. C.L. assisted with MOCVD. T.C. wrote the manuscript with help from Z.N. All authors have given approval to the final version of the manuscript.

Notes

The authors declare no competing financial interest.

ACKNOWLEDGEMENTS

This work was primarily supported by the Air Force Office of Scientific Research (grant no. FA9550-21-1-0323) and the MURI project (grant no. FA9550-18-1-0480). Additional funding was provided by the Office of Naval Research (grant no. N000142212841) and the Samsung Advanced Institute of Technology. T.C. acknowledges support from the Kadanoff-Rice Postdoctoral Fellowship through the National Science Foundation (NSF) Materials Research Science and Engineering Center (MRSEC) under grant no. DMR-2011854. F.M. acknowledges funding from the NSF Graduate Research Fellowship Program under grant no. DGE-1746045. Research conducted at the Center for Nanoscale Materials, a U.S. Department of Energy (DOE) Office of Science User Facility, was supported by the DOE Office of Basic Energy Sciences under contract no. DE-AC02-06CH11357. A.R. and the STEM imaging at the Cornell Center for Materials Research were supported by the NSF MRSEC program under grant no. DMR-1719875. The Thermo Fisher Spectra 300 X-CFEG was acquired with support from PARADIM, an NSF Materials Innovation Platform (grant no. DMR-2039380), and Cornell University.